# Seeing Beyond the Surface:
# Understanding and Tracking Fraudulent Cyber Activities

[1]Longe,O. B.  [2]Mbarika, V.
*Int. Centre for IT & Development*
*Southern University*
*Baton Rouge, LA 70813*

[3]Kourouma, M
*Dept. of Computer Science*
*Southern University*
*Baton Rouge, LA 70813*

[4]Wada, F.  & [5]Isabalija, R
*Nelson Mandela School of Public Policy*
*Southern University*
*Baton Rouge, LA 70813*

**Abstract**
The malaise of electronic spam mail that solicit illicit partnership using bogus business proposals (popularly called 419 mails) remained unabated on the internet despite concerted efforts. In addition to these are the emergence and prevalence of phishing scams that use social engineering tactics to obtain online access codes such as credit card number, ATM pin numbers, bank account details, social security number and other personal information[22]. In an age where dependence on electronic transaction is on the increase, the web security community will have to devise more pragmatic measures to make the cyberspace safe from these demeaning ills. Understanding the perpetrators of internet crimes and their mode of operation is a basis for any meaningful effort towards stemming these crimes. This paper discusses the nature of the criminals engaged in fraudulent cyberspace activities with special emphasis on the Nigeria 419 scam mails. Based on a qualitative analysis and experiments to trace the source of electronic spam and phishing e-mails received over a six months period, we provide information about the scammers' personalities, motivation, methodologies and victims. We posited that popular e-mail clients are deficient in the provision of effective mechanisms that can aid users in identifying fraud mails and protect them against phishing attacks. We demonstrate, using state of the art techniques, how users can detect and avoid fraudulent e-mails and conclude by making appropriate recommendations based on our findings.

**Keyword**: *Spammers, Scamming, E-mail, Fraud, Phishing, Nigeria, IPLocator*

## I.  INTRODUCTION

Cybercrime refer to misconducts in the cyber space as well as wrongful use of the internet for criminal purposes. Various categories of these crimes include cyber stalking, phishing (identity theft), virus attacks, malware attack, the use of anonymous proxies to masquerade and sniff information and the popular electronic spam mail problem. Unfortunately, Cybercrime seems to be yielding much to criminals all over the world so the malaise is not going tom be curbed without some resistance from the criminals.  Offline crime rates have reduced because the offline criminals have gone cyber. In fact, it is highly likely that Cybercrime and its perpetrators will continue developing and upgrading knowledge to stay steps ahead of the law.  It is generally believed that most fraudulent mails in cyberspace and phishing attacks originate from or are traceable to Nigeria or Nigerians in other nations.

These assumptions remains to be validated using empirical research [27]. Since the implication of cybercrime is well beyond its immediate points of perpetration and utilization of online facilities depends strongly on consumer trust, attention must be given to factors that can impede the effective use of information technology platforms for electronic transactions. The negative effects of online criminal activities on the use of the internet for electronic commerce, e-banking and other forms of usage has therefore increased interest in studying the factors that motivates these criminals, their tactics and what can be done to mitigate their activities[13][17].

The remaining part of the paper is organized as follows: In section 2 we discussed cybercrime in Nigeria with particular emphasis on scamming and phishing. Section 3 examined the nature of the cyber criminals, the victims and the tools used by these criminals. The scammers mode of operations are discussed in section 4. In section five we provide simple guidance on how users can trace the source of suspected scam mails. We concluded the paper in section 6 by suggesting possible measures that can be adopted to address the problem and gave direction for future research.

## II. CYBERCRIME IN NIGERIA

Cybercrimes share some similarities with crimes that have existed for centuries before the advent of the cyber space. The major difference is that the internet now provides an electronic platform with the advantages of speed, anonymity and a tool which increases their potential pool of victims [1] Among the numerous crimes committed daily on the Internet, Nigeria and some other nations on the West African coast are reputed to be at the forefront of sending fraudulent and bogus financial proposals all over the world. The damaging implications resultant on the image of the Nigerian nation and the negative impact this trend has had on e-mail infrastructures are clearly evident. The United States Internet Crime Complaint Centre defined the Nigerian scam as "Any scam that involves an unsolicited email message, purportedly from Nigeria or another African nation, in which the sender promises a large sum of money to the recipient. In return the recipient is asked to pay an advance fee or provide identity, credit card or bank account information[25]. While this genre of cyber crimes are generally targeted at individuals, they require a degree of ingenuity to dealt real damage on the victims. The damage done manifests itself in the real world as human weaknesses such as greed and gullibility are generally exploited.

No discussion about organized financial crime is complete in Nigeria without mentioning Fred Ajudua a Nigerian fraud kingpin who is currently standing trial in Nigeria for alleged involvement in advance fee fraud "419" charges. These charges relate to collecting money under false pretences from Nelson Allen, a Canadian who allegedly lost $285,000 to Ajudua and is the only foreigner to have given mail-fraud evidence in a Nigerian court of law. Other suspected Ajudua victims include Technex Import and Export Company of Germany, who lost the equivalent of $230,000 USD, and a German woman who lost the equivalent of $350,000 USD trying to collect purported dividends left by her late husband. It is generally believed that there are many more victims yet to come forward [28].

### A. Scamming

Scamming or 419 fraud is usually in the form of "Advance Fee Fraud" (named after the relevant section of the Criminal Code of Nigeria that deals with such crimes). It begins when the target receives an unsolicited fax, e-mail, or letter often concerning Nigeria or another African nation containing either a money laundering or other illegal proposal. With the increase in use of internet facilities for electronic commerce, online banking, social networking and other financial transactions phishing attacks are also on the increase. Spamming was said to be one of the most prevalent activities on the Nigerian Internet landscape accounting for 18% of all online activities amongst others [11]. Information released by the United States Internet Crime Complaint Centre between 2006 and 2008 brings the Nigerian Spam situation to the fore as the nation maintained a high position among the first ten nations that serve as the source of Spam all over the world (see tables below). The report showed that confidence fraud, computer fraud, check fraud, and Nigerian letter fraud round out the top seven categories of complaints referred to law enforcement during the year. Of those complaints reporting a dollar loss, the highest median losses were found among check fraud ($3,000), confidence fraud ($2,000), Nigerian (west African, 419, Advance Fee) letter fraud ($1,650). Diplomatic missions around the world warn visitors to various West African countries such as Nigeria, Côte d'Ivoire, Togo, Senegal, Ghana, Burkina Faso and Benin Republic of susceptibility to 419 scams. Countries outside of West Africa with 419 warnings are South Africa, Spain, and The Netherlands.

The effect of fraudulent spamming activities can be measured by the pressure the volume of spam messages placed on internet bandwidth, thus slowing it down. This is not helpful in an age when subscribers are clamoring for faster connections. It also increases the dial-up costs by extending the time a person spends reviewing e-mail. When Spammers use false e-mail addresses and users attempt to respond to them, the e-mail bounces around in cyberspace loops creating huge administrative loads for Internet infrastructures. Other hidden costs involve the claims made on two precious human resources: time and energy. Computer users can spend hours attempting to identify the original sender of an e-mail.

.

**Table 1: Amount Lost by Selected Fraud Type for Reported Monetary Loss**

| Complaint Type | % of Reported Total Dollar Loss | Of those who reported a loss the Average (median) $ Loss per Complaint |
|---|---|---|
| Nigerian Letter Fraud | 1.7% | $5,100.00 |
| Check Fraud | 11.1% | $3,744.00 |
| Investment Fraud | 4.0% | $2,694.99 |
| Confidence Fraud | 4.5% | $2400.00 |
| Auction Fraud | 33.0% | $602.50 |
| Non-delivery | 28.1% | $585.00 |
| Credit/debit Card Fraud | 3.6% | $427.50 |

*Source Internet Crime Complaint Centre Report (2006-2008) [25]*

**Table 2: Top Ten Countries - Perpetrator of Cybercrime**

| Year 2006 | | Year 2007 | | Year 2008 | |
|---|---|---|---|---|---|
| United States | 60.9% | United States | 63.2% | United States | 66.1% |
| United Kingdom | 15.9% | United Kingdom | 15.3% | United Kingdom | 10.5% |
| Nigeria | 5.9% | Nigeria | 5.7% | Nigeria | 7.5% |
| Canada | 5.6% | Canada | 5.6% | Canada | 3.1% |
| Romania | 1.6% | Romania | 1.5% | China | 1.6% |
| Italy | 1.2% | Italy | 1.3% | South Africa | 0.7% |
| Netherlands | 1.2% | Spain | 0.9% | Ghana | 0.6% |
| Russia | 1.1% | South Africa | 0.9% | Spain | 0.6% |
| Germany | 0.7% | Russia | 0.8% | Italy | 0.5% |
| South Africa | 0.6% | Ghana | 0.7% | Romania | 0.5% |

*Source Internet Crime Complaint Centre Report (2006-2008) [25]*

With volumes such as this, tremendous burdens are shifted to the ISP to process and store unnecessarily large amount of data [29]. Users bear the costs involved with Spam. Payment occurs when a user is "taken" by the Spammer and pays for services or products never to be realized. This happens more frequently than one might expect. In [30] it was opined that there are also the hidden costs masked as "yearly access fee increases" to help the ISP provide better service to the users. This refers to dealing with regular disruptions to the integrity of the systems which results into the ISPs simply passing their costs down to subscribers and other stakeholders. This has obviously contributed to many of the access, speed and reliability problems seen with lots of ISPs today. Spamming and phishing interferes with ordinary e-mail communication, reduces employee productivity and engender lack of trust in the electronic infrastructure. It has been projected that spamming will cost 1.4 percent of employee productivity, or N131,100 per year per employee in Nigeria, the equivalence of $874 in the US [11]. Filtering Spam is therefore an extremely urgent problem. With the escalation in the volume of spam mails on the webscape, organizations are also being subject to mounting pressure to deal with issues regarding employee productivity, morale, sexual harassment and congestion of the e-mail infrastructure.

*B. Phishing*

Phishing is a social engineering scam that involves luring unsuspecting users to take a cyber-bait much the same way conventional fishing involves luring a fish using a bait. Phishing deceives consumers into disclosing their personal and financial data, such as secret access data or credit card or bank account numbers, it is an identity theft. It is an attempt to elicit a specific response to

.

a social situation the perpetrator has engineered (Teri, 2002). Identity theft schemes take numerous forms and may be conducted by e-mail (phishing), standard mail, telephone or fax. Thieves may also go through trash looking for discarded tax returns, bank records, credit card receipts or other records that contain personal and financial information so as to use someone's personal data to steal his or her financial accounts(http://www.irs.ustreas.gov)

Phishing scams are on the increase in Nigeria. The most recent phishing attacks were on the customers of Interswitch, the banking and financial system backbone organization with the highest customer base in electronic transactions in the country. According to APWG, The number of unique phishing websites detected during the second half of 2008 saw a constant increase from July – October with a high of 27,739 (http://www.antiphishing.org). The Nigeria Deposit Insurance Corporation (NDIC) disclosed in its 2007 annual report and statement of account that under-hand deals by bank staff, among others, resulted in attempted fraud cases totaling N10.01 billion and actual losses of N2.76 billion in 2007 [15]. With the present economic downturn and appropriate technology, fraudulent actions are most likely to increase and phishing remains one of the means of committing "fraudulent crimes without borders". The case of three people, two of them Nigerians, who were arrested by the police after having conducting a phishing scheme with approximately 30 victims was reported in [2]. They posted fake e-mails to the clients of a local bank in India asking them to visit a link which required them to enter private details such as credit card number, PIN and other information. Once the users entered the information, the phishers received and used it to transfer over $100,000 from the victims accounts.

In their paper "why phishing works" Rachna et al [20] came up with the fact that good phishing websites fooled 90% of users and existing anti-phishing browsing cues are ineffective. It was also reported that 23% of users do not look at the address bar, status bar, or the security indicators and on the average users made mistakes 40% of the time in identifying phishing websites. Phishing attacks have convinced up to 5% of their recipients to provide sensitive information to spoofed websites. By hijacking the trusted brands of well-known institutions, phishers are able to convince a small percentage of recipients to respond to them[7]. The question however remains as to whether or not the success of phishing scams are a result of underlying security flaws in web security or simply the result of laxities in user assessment of phishing offers.

## III. THE CYBERCRIMINALS, THEIR VICTIMS AND THE TOOLS.

All forms of crime involve two major players are involved. The criminals and the victims. To analyze conventional criminal conducts, sociologists focus on two main issues viz the crimes and the criminal committing the crime [4][6]. Criminologists are primarily concerned with the sociological factors that cause, or are correlated with a person becoming a criminal and engaging in criminal activities. The social learning theory [5], social bonding theory [12 and rational choice theory are some of the sociological theories that address these issues. Rational choice theory argues that people make a basic decision to commit a crime, or to not commit a crime, based on a simple cost-benefit analysis. The rational choice theory focused on non-sociological factors that can influence the decision to commit crime [31]. For example, electronic mechanisms such as user ID, automated access control systems and surveillance camera can serve as deterrents because they increased the perceived risk of being apprehended [6]. To effectively deal with cybercrime, an understanding of the crime, the criminals, the victims and the tools are required.

### A. The Criminals – The Nigerian 419 Spammers

To understand the nature of cyber criminals, it is important to have an idea of the general misconceptions internet users have about the criminals. In the article "Piercing the darkness – Misconceptions about Cyber criminals" Thinkquest [23] identified four major misconceptions about cybercriminals. These are:

(1) Misconception 1: All cyber criminals are smart but social misfits
(2) Misconception 2: Cyber criminals are not "real" criminals
(3) Misconception 3; Teenagers with computers are all cyber criminals
(3) Misconception 3: All cyber criminals have the same characteristics

Some studies also opined that the 419 cohorts are illiterates or semi-illiterates. These conclusions were based on the lack of fluency in the use of English, grammatical errors, lexis and the general syntax of the sentences in most 419 mails [21].

.

Other studies claimed that the fraud team consists of Nigerians in Diaspora operating (with assistance from some confederates in the West African coasts) from various places in Europe, Asia, America, Australia and the middle east [19]. There are people who also believed there are not enough computers in Nigeria for these criminals to perpetrate scams on a scale this large and that scammers are all from a particular tribe in Nigeria. While some of these assumptions are tenable others are clearly misguided.

The Nigerian Spammers are a mixture of criminally minded elites and semi-elites who simply moved from conventional offline 419 schemes to an electronic platform offered by the Internet. It is clearly an extension of the popular fax and postal mails scams that originated in Lagos, Nigeria back in the 80s[10]. Qualified graduates and undergraduates from universities and polytechnics, high school and secondary school students, kingpins of the underworld, semi-illiterates and the unemployed are all involved in these heinous crimes. The level of co-ordination and sophistication now involve very high degree of intelligence and software instruments that are difficult to beat by spam filters and human beings alike. The cybercrime trade is fast turning to an internationally co-ordinated and controlled system. There are indications that computing and allied professionals are recruited specifically to provide expertise in the context of hacking and designing tools that will assist in beating security measures.

The evolution of fixed wireless facilities providing internet access anywhere and anytime has enabled a migration of these criminals from the public internet access points to the comfort of their homes and offices. There is therefore a very high degree of mobility of criminals and the network is further enhanced by the availability of high speed wireless connections and the Global System for Mobile communication (GSM). Today, there are in existence, very organized cartel-like organizations mushrooming in different location, especially in the south west and south eastern part of the country. A common impetus among these cyber criminals is their desire to make "big money" that will enable them afford buying "big cars" and "powerful camera phones". These criminals have successfully swindled locals and foreigners of their money and these success are measureable in terms of very fat bank accounts and exotic cars seen among youths involved in scamming. Their success serves as an impetus for others to get involved and wait for their "lucky day".

## B. The Victims

Victims of cybercrime are many and varied. They range from individuals, business organizations, religious organizations, philanthropic organizations and educational institutions. Understanding who the criminal is likely to target can assist in taking preemptive actions to forewarn and prepare for all forms of attack. Intelligent criminals always target people whose circumstances are loaded with some forms of vulnerability. The yahoo boys are fond of people who are easy to deceive. Oral interviews of cyber criminals in cyber cafes in Nigeria yielded the response that "yahoo-yahoo business is all about deceit, if you are gullible, then you become a victim". Users of internet facilities have to be on guard against all forms of solicitation that comes from strangers with very enticing dividends. One nature that the cybercriminal prey upon most is gullibility and greediness. There are users who see the internet as a very easy tool with which to make money. They gullibly provide private information and account details. Usually, they have some dividends from such transactions at the initial stage. This success takes them deeper into the con as they build more relationship with the scammers and become embroiled in legal and financial entanglements out of which only the perpetrator will make profits.

Others are enticed by advertisements and offers that invite them to try out new products and means of making money. The Diversity Visa Lottery spam is used to fool local victims who are desperate about travelling abroad. Foreigners are engaged and embroiled in conference invitations and asked to pay registration fees that are cornered by the con men. Fictitious websites are set up for employment purposes using phishing. In some cases, these criminals also set up dating sites requesting for personal information and luring the victim to play along as they buy time to extract credit card numbers and render the customer bankrupt. Unfortunately the sheer embarrassment of being defrauded online has prevented most victim from reporting their ordeal, preferring to silently bear the pain and possibly learn their lessons in a bitter way. It is unbelievable at times to imagine the extent of activities involved between the criminal and the victim before money exchange hands. Some correspond for weeks and months, planning, scheming together and building trust with someone they have never met and whose credentials they cannot prove. Victims are enticed by stories that are usually emotion-laden, financially-promising or religiously-toned. The most unfortunate thing about it all is that most victims act alone. No consultation or efforts are made to seek advice from experts or security agent before committing their time, money and energy to deals offered on the internet.

*C. The Scammers Tools*

A combination of social engineering and programming skills are the most potent tools in the hands of the 419 scammers. In order to reach a large volume of users, the scammers require an equally large number of email addresses. These are usually collected by using programs known as spam-bots to search for email addresses listed on web sites and message boards, by performing a dictionary attack (pairing randomly generated usernames with known domain names to 'guess' a correct address) or by purchasing address lists from individuals or organizations. Once they have addresses, spammers can use programs known as "bulk mailers" to automate the sending of spam. These programs can send huge volumes of email messages in a small amount of time. Some bulk mailing programs engaged by the spammers use open-relays to send messages, effectively hiding the true address of the spammer. Bulk mailers can also fabricate the *from* address in email message headers to further hide the identity of the spammer [8]. Another technique spammers utilize to send emails is with the use of *zombie network*s, also known as *bot network*s. Zombie is the term given to a computer that has been infected by a virus, worm, or Trojan Horse [9], which allows remote entities to take control and use it for their own (usually illegal) purposes. A large amount of these computers, usually called a *network* or *army* can be co-opted to send spam emails, requiring little of the spammer's own computing power and network bandwidth. This technique is also popular as it protects the identity of the spammer [18].

Another popular method employed by scammers is the use of dating sites as a powerful tool to get attention and e-mail addresses. A number of victims have fallen victim to dating scams. Religious persuasions and emotional-laden mails are designed to attract attention and sympathy from religious organizations. Just as the web security community developed personalized electronic mail filtering systems, scammers also develop tactics that are personalized. They profile individuals, trace their business history with individuals they have been involved with in the past. The knowledge of old business acquaintances abroad are employed to compose emotion-laden letters with bogus business proposals from

Nigerians purportedly in government looking for opportunities to launder money abroad through a friend or two. Keystroke loggers are also used by these criminals to carefully collect personal information from unsuspecting victims. This trick is employed when unsuspecting users log on to the wrong website during a request for program update. This is particularly targeted at financial organizations. Most phishing attacks from either state that there has been some sort of fraud detected on bank accounts or that for security reasons the company just wants everyone to validate their usernames and passwords. In either event, the attack preys on fear and naiveté to get people to respond by providing sensitive information such as usernames, passwords, account numbers, etc. Cyber criminals can combine phishing with the 'Nigerian Bank Scam' to use greed rather than fear as the driving force to prey on individuals [24].

## IV. THE MODUS OPERANDI

In order to understand current techniques used by the 419 scammers, we set up electronic mail accounts on popular free e-mail clients such as yahoo, hotmail and excite for a six month period. Additionally, we also operated the e-mail account from locations outside Nigeria, specifically in the United States of America and United Kingdom to ascertain if there are variations in the content and style of electronic mails received from the conmen at these locations. We deliberately did not subscribe to any form of promotion, materials or subject while registering the e-mail account. This is to enable us have a clean slate to operate and avoid other forms of spam. We employed the software tools IP2Location, IPGeolocator and GeoBytes IP Locator to track the source of the electronic mails that were received. We provide samples of these e-mails in the figures below. A breakdown and analysis of the various genre of e-mail is presented in Table 3.

.

### Table 3: Genre of Spam Mails and their Statistics

| Type | Spamming Technique Employed | Total No of Mails | Total of False | % of false Negative |
|---|---|---|---|---|
| **One line Subject Link** | One sentence e-mail consisting only a weblink that take the recipients to another page where the actual scam is to be perpetrated | 78 | 21 | 26.9% |
| **Spoofing/phishing** | This type of mail comes with fictitious sender addresses. They usually direct users to respond to an entirely different address(es) from the sender address. Usually, the two addresses looks similar. i.e longeolumide@yahoo.com and olumidelonge@yahoo.com | 187 | 73 | 39.0% |
| Image With Superimposed Spam | These e-mails contains an image of any company or organization on which text are superimposed to communicate to the user. An example is shown in Fig. 2 | 152 | 62 | 40.7% |
| Subject Title and Sender ID Similarity | This e-mail type makes the subject of the e-mail and the sender header the same.  sender and specify the same. i.e a  message title: From Longe Sender : From Longe | 107 | 24 | 22.4% |
| Bayesian Poisoning | This type of e-mail usually contain a link to another website with an e-mail body loaded with a lot of grammatical error and meaningless sentences to confuse the Bayesian filter. At other times, tokens are manipulated using word toggling, a combination of numerals and letters i.e AgeiNg numbe3rs etc. | 261 | 77 | 29.5% |
| Attachment Only E-mail | These e-mails contain just an attachment or two with no contents at all. | 78 | 19 | 24.3% |

*Note: False negatives are spam mails wrongly classified as good mails (found in the Inbox)*

As at the time of writing, a total of 1113 unsolicited electronic mails have been received from the e-mail accounts set up for the research between April – October, 2009. Out of this number, purely pornographic e-mail numbered 45 with 37 of them received when the client moved between Europe and America.  Marketing e-mail for medication, commercial products and webinars were 76 in number. The remaining 955 mails collated from all the various locations are purely 419 mails. Out of this number, 187 are purely spoofed/phishing e-mails that direct users or recipients to websites set up for collecting personal information. Others in this category direct the reader to subscribe to one product or the other with the clause that "if the e-mail is received in error, the reader should click a link to unsubscribe" in order to stop receiving the mail. A click on such link is a subtle way to validate e-mail addresses as it reflects the fact that the mail has been received by a user. There is also a degree of overlap among 68 of the remaining e-mails making them fall into other categories.

These were discarded. Of the 800 remaining 419 e-mails, 633 were selected because they  satisfy stratification for language, type and content.

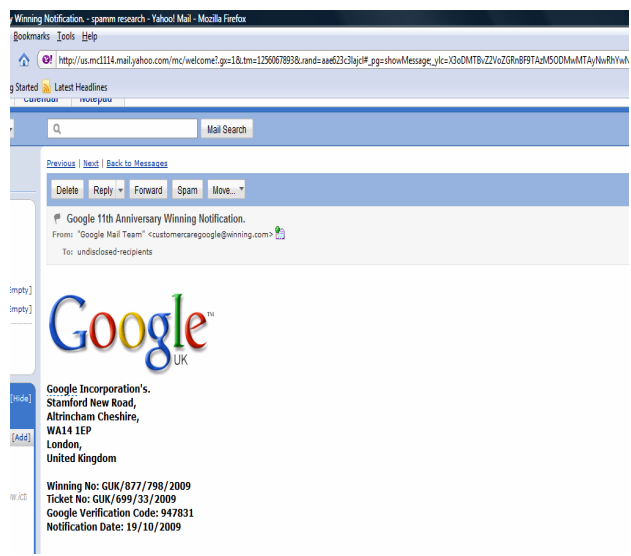

*Fig. 1: Lottery Winning and Financial Scam*

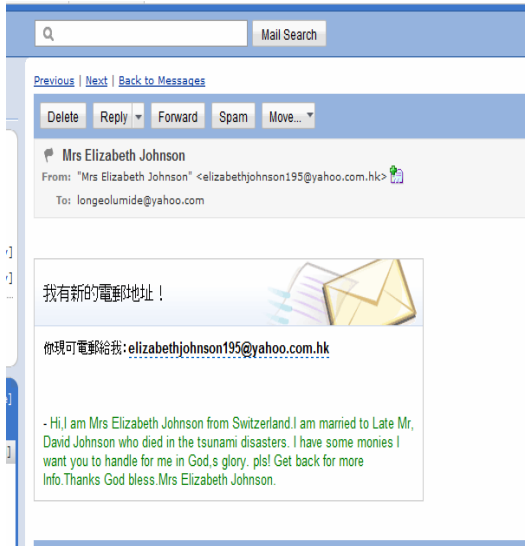

*Fig. 2: E-mail Scam using Images*

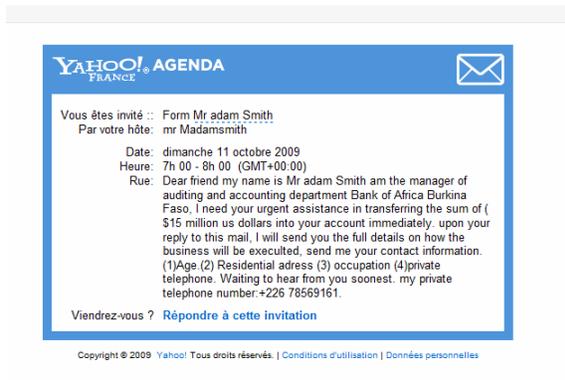

*Fig 3. Images Spam With Text Superimposed.*

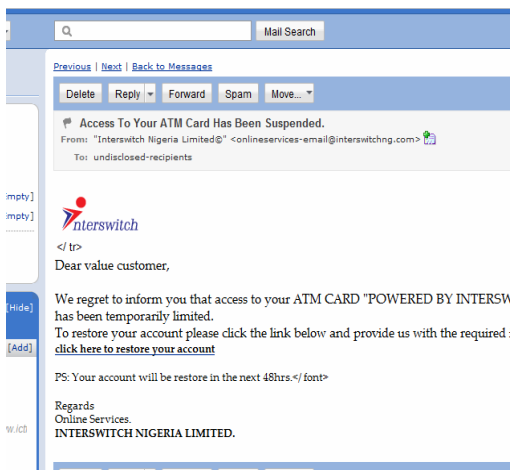

*Fig 4: Pure 419 Scam for Phishing*

## V. EXPERIMENT WITH E-MAIL SOURCE AND IP ADDRESS LOCATION

It is important as a first line of defence for e-mailers to know how to track e-mail origin. We use yahoo mail account for our demonstration here since it is one of the most popular e-mail clients. The sequence is almost the same for all other e-mail clients. The User need to be able to view the full header of an e-mail before they can track the origin of any mail. For yahoo, the "Full Header" option can be seen at the bottom right side of each e-mail message received. The header section contains a lot of information relating to the mail. Sometimes there are several **Received From's** in the header. This is so because the header contains the IP addresses of all servers involved in routing e-mail from one point to another. With this addresses a user can track an e-mail origin and possibly the identity of the sender.

We selected some mails randomly from our corpus of spam mails. To find the actual location from which the e-mail originates we pick the "Received From" IP that is at the bottom of the list on the header view. The results below is among others for an e-mail purportedly originating from Interswitch Nigeria limited claiming that **"an account has been suspended and personal details should be provided for reactivation".** For the sake of brevity, a portion of the header details is given below (Fig. 3).

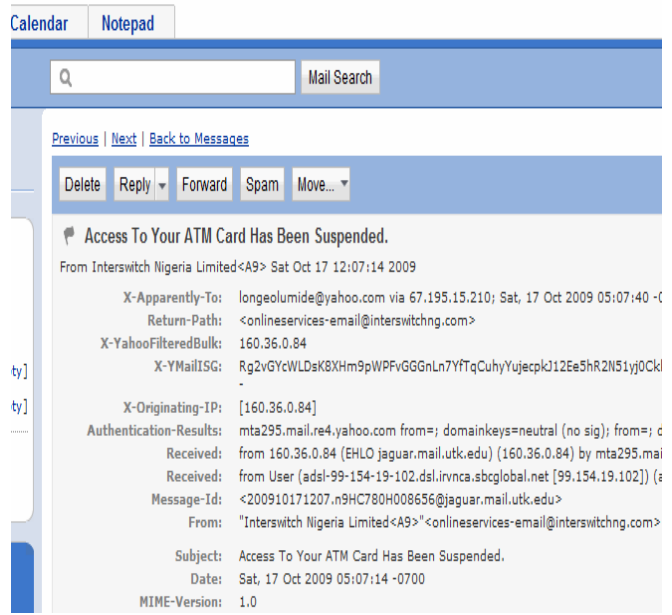

*Fig. 5:  Letter Purportedly from Interswitch Nigeria Limited*

.

The next step is to look up the IP address. There are freeware open source software that are designed to do just that. In this work, we use the **IP2Location** and **GeoBytes IP Locator** and **TraceE-mail**.

We checked for the source of this e-mail using the IP address on the e-mail as well as the e-mail address itself. **IP2Location** produced the result below when fed with the IP address.

Now, you can run IP address query in your desktop even without network. Please download and install the IP2Location Application today.

| IP Address | Country (Short) | Country (Full) | Flag | Region | City | ISP | Map |
|---|---|---|---|---|---|---|---|
| 160.36.0.84 | US | UNITED STATES |  | TENNESSEE | KNOXVILLE | UNIVERSITY OF TENNESSEE | MAP IT! |
| 160.36.0.84 | US | UNITED STATES |  | TENNESSEE | KNOXVILLE | UNIVERSITY OF TENNESSEE | MAP IT! |
| 160.36.0.84 | US | UNITED STATES |  | TENNESSEE | KNOXVILLE | UNIVERSITY OF TENNESSEE | MAP IT! |

*Fig 6: Result from IP2Location on the validity of the IP address*

To further substantiate this result, we used another locator **IP Location Finder** and **IP Locator** and obtain similar results shown below.

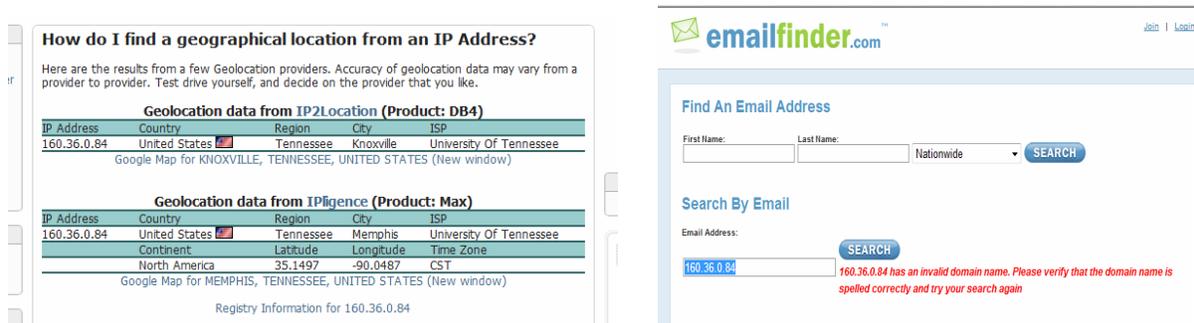

*Fig.7: Result from IP Location Finder and IP Locator on the validity of the IP address*

Next we use the e-mail address "onlineservices-email@interswitchng.com" to check the validity of the e-mail on IP Locator and E-mail trace. We obtained the result showing the letter emanates from Tennessee in the United States.

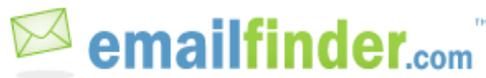

**valid Domain Search**

The email address onlineservices-email@interswitchng.com appears to be invalid. Click here to run another search.

If you received a message from this address, the email was likely "spoofed" and actually from someone else. This is quite common.

*Fig. 8; Result from IP Locator on the validity of the e-mail address*

.

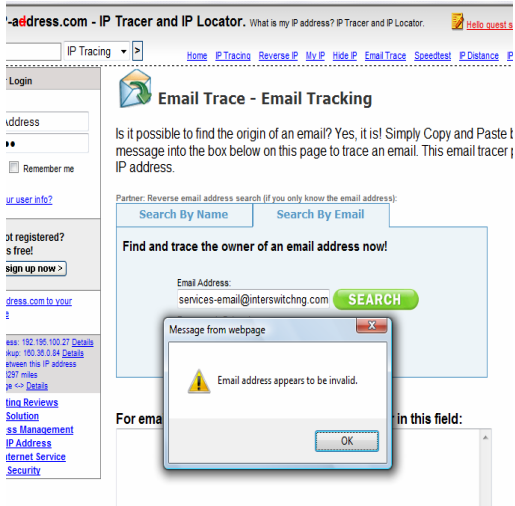

For a letter purportedly originating in Dubai (68.142.236.201) and inviting the authors for a job interview in Dubai. The test showed that the letter emanated from Sunnyvale, United States

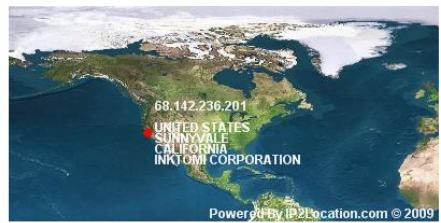

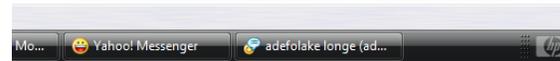

*Fig. 9: E-Mail trace validity of the e-mail address*

For a letter purportedly posted inviting us to a conference from "Miss (Regina Pedro) presently working with (GLOBAL YOUTH ORGANIZATION FOR HUMAN RIGHT) California, USA" the e-mail address finder produced the result below.

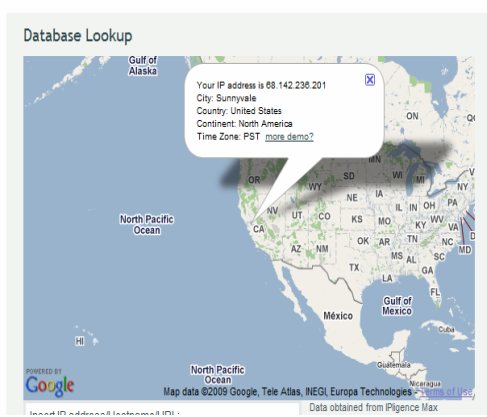

*Fig. 10: I P Goe Locator Validation*

*Fig. 11: Result from IP2Location With Map*

To be sure that the locators are reporting correctly, we search for the location of the computer being used at the time of writing this paper using IP2Locator and IP tracer. The results obtained are shown in Fig. 12. The authors were actually wordprocessing the paper at the Southern University, Baton Rouge, Louisiana, USA

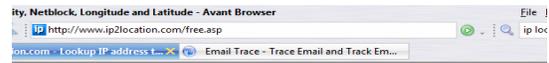

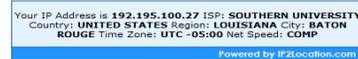

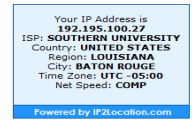

*Fig. 12: Authors Location Authoring Paper*

## VI. DISCUSSION OF FINDINGS

For the purpose of our research, we used medium spam filtering for the e-mail accounts by setting the spam filtering option to medium. This is intended to balance the degree of indirection between expert users and novice when using e-mail accounts. Our findings from the analysis of the mail corpus showed that the greatest challenge the scammer face is how to beat existing filtering systems. As a result of this, they craft their e-mails in such a way that the victim get baited to phishing websites where they can divulge information from which the scammer can make financial gains. They therefore stick to tactics yielding dividends until the security system evolve to tackle it. Interestingly, we received mails from mailboxes set up abroad informing us about access to ATM CARD "POWERED BY INTERSWITCH" being temporarily limited" we were asked to click a link to restore the account in 48 hours. Interswitch, is the backbone organization that links all financial institution on ATM and smartcards in Nigeria. The link takes us to *http://ussonline.free.fr/uss/administrator/compon ents/com_remository/login.html.* The letter itself purportedly originated from onlineservices-email@interswitchng.com>.

The image–based spam with super-imposed text/content is a scheme that is yielding very positive results for the scammer. Over 40% of such mails get past current filters. This is followed by spoofing and Bayesian poisoning. A lot of efforts in literature has focused on dealing with Bayesian poisoning, unfortunately, it seems the spammers are always one step ahead as there a number of ways a mail can be composed to defeat these content-based spamming method. Lately, most Bayesian-poisoned e-mails are coined to guide users to phishing websites. The percentage of such e-mails that beat current spam filters are outstanding. The scammers tactics to also use attachment only, one subject link and sender and subject similarity are producing commensurate results (over 24% on the average). The spammer needs a turnaround of less than 5% to get good returns for their efforts.

This research however cast some shadows on the general assumption that most fraudulent mails originate from Nigeria. While not excusing this view, empirical research is warranted to study trends as far as spam mail sources or origins are concerned. Most of the e-mails passed through the locators or IP trackers yield results that showed that they emanated from places outside Nigeria.

The interesting thing is that some of them lay claim to wanting to solve problems for people living in Nigeria. The validation of previous claims about the sources of fraudulent e-mails will have to be subjected to scientific experiments such as we have done in order to direct filtering efforts and legislation to reach out to all instances of location of the spam problem.

## VII. CONCLUSION

For spamming and identity theft to be successful weaknesses such as ignorance, oversight and lack of awareness are exploited by the spammers. Most users are devoid of the knowledge of the workings of the mailing system as well as online vulnerabilities. For users generally, disposing off bank statement carelessly close to ATM machines calls for great concern. Customers are unconscious of the fact that scammers can fiddle with trash bins to look for account details and use them to withdraw funds from their account. It is worth noting that phishers are getting smarter and it will not be out of place to say that in the future their methodologies could advance to using some elements of frustration and disruptions to financial network systems to compel users to provide information. A qualitative interview of regular e-mailers in the UK, the US and Nigeria on the problem of spam mails revealed that some users are not aware that there are provisions within the e-mail client to mark messages as spam in order to prevent them from receiving spam from the same source in the future.

The awareness of blacklisting spam mails and whitelisting wrongly classified mails is also low. Most users are not conscious of or have never use reporting facilities that help authorities track phishing websites. An average user cannot differentiate between a legitimate Universal Resource locator (URL) and a fake. In some cases, users are not knowledgeable about how to identify and distinguish between good and fake security indicators on visited websites. This lack of awareness and information on internet scam and what to look out for in order to prevent phishing attacks is one of the major reasons why phishing and other electronic fraud are successful.

### A. Recommendations

Considering the far reaching effects of scamming, it is appropriate at this juncture to recommend that service providers and organizations orient users

.

about their vulnerabilities online and be equipped with standard practice measures to forestall and prevent them from falling victim to cybercrime. The present web design security architecture and common e-mail clients are deficient in the provision of effective mechanisms that can aid users (as the last line of defense) to identify and avoid fraud mails and phishing attacks. It is important to address the issue of security indicators in web design. These indicators have to be made more glaring and obvious enough for users to identify. E-mail clients can also make it easy for users to avoid and fight spamming by providing full lexical information in the headers rather than the codification of IP addresses and Mailing addresses that we currently have. [17]

*B. Direction For Future Work*

In the future, we intend to perform a comprehensive experiment relative to the source/origin and location on the corpus of spam e-mails and phishing websites that we are building. This will provide a scientific basis regarding the type of e-mails that come from particular origins and show the skewness of these mails by location.

## ACKNOWLEDGEMENT

This work is supported by the (2009) University of Ibadan MacArthur Foundation Grant for Manpower Training & Development.

.